# Cognifying Education: Mapping AI's transformative role in emotional, creative, and collaborative learning

**Mikael Gorsky[1], Ilya Levin[2]**

[1]*Holon Institute of Technology, Holon, Israel, MikaelG@hit.ac.il*

[2]*Holon Institute of Technology, Holon, Israel, LevinI@hit.ac.il*

## Abstract

*Artificial intelligence (AI) is changing educational practice and testing old assumptions about teaching and learning. This article combines ideas from three recent books with evidence from popular AI podcasts, published research, and public releases from Anthropic. The books are Genesis by Kissinger, Schmidt, and Mundie, Co-Intelligence by Mollick, and The Inevitable by Kelly. We examine seven domains: emotional support, creativity, contextual understanding, student engagement, problem-solving, ethics and morality, and collaboration. For each domain, we describe what AI can do, where it can change practice, and which classroom practices are starting to work. We give equal weight to theoretical analysis and to real-world observation. We find that thoughtful use of AI can support human educators and enrich learning in its cognitive, social, and emotional dimensions. Our outlook is optimistic and responsible. Educators and students should take an active part in shaping how AI enters education. Their aim should be to strengthen human creativity, ethical reasoning, and collaboration, and to keep human values at the center.*

## 1. Introduction

The rapid growth of AI capability has started an intense debate in education. Large language models such as GPT-4 and Claude are the clearest examples of this growth. Early public opinion often treats AI in one of two ways. Some people see it as a threat to the teacher's unique role. Others see it as a tool that can only automate routine tasks. A closer look shows a more complex picture. With proper guidance, AI systems can extend the reach of educators and enrich student learning in new ways. We identify seven areas where AI can have a significant impact: (1) emotional support, (2) creativity, (3) contextual understanding, (4) student engagement, (5) problem-solving, (6) ethics and morality, and (7) collaboration. These seven themes give our inquiry its structure.

Recent books by prominent thinkers offer frameworks for understanding how AI may change human roles and capabilities. Kissinger, Schmidt, and Mundie (2024) place AI's impact on human knowledge and agency in a broad geopolitical and philosophical context [1]. Mollick (2024) focuses on practical collaboration between people and AI in work and education. He describes AI as a tutor, a coach, and a creative partner [2]. Kelly (2016) presents AI as an inevitable force that will "cognify" every part of life [3]. He argues that people must learn to work alongside intelligent machines.

Popular AI podcasts and public content from AI research labs add real-world evidence. The podcasts include *Hard Fork*, *The Ezra Klein Show*, *Latent Space*, and *The Cognitive Revolution*. On these programs, teachers and technologists describe how classrooms already use AI. They also describe the problems and successes that are appearing. On *Hard Fork*, for example, Wharton professor Ethan Mollick explained how he brought generative AI into his teaching. He requires his students to use AI for certain assignments [4]. Such firsthand accounts add to data-driven reports, such as Anthropic's large-scale study of how students use AI [5]. They also add to Anthropic's research on the values that AI models express in real conversations [6]. Together, these sources give an exploratory and optimistic view of AI's potential in education. That potential depends on careful and ethical integration.

The seven themes are examined in turn below. For each domain, we compare common first assumptions with what AI-supported education looks like in practice. We draw on the arguments of leading thinkers, on findings from research and practice, and on examples of AI used for deeper learning. Our tone is optimistic. We highlight opportunities and effective strategies, and we also note the need for continued vigilance, especially on ethics and equity. The Conclusion brings these insights together and discusses how educators can actively shape an AI-enhanced future of learning.

## 2. AI and Emotional Support

Many people believe that only human teachers can give students empathy and emotional support. In this view, AI would be "cold" and unable to understand feelings. Emotional intelligence and genuine empathy are indeed deeply human traits. Early implementations and studies, however, suggest that AI can simulate empathic support in ways that help students. The AI itself has no consciousness or emotions. It can still detect emotional states by analyzing text, tone of voice, and even facial expressions, and it can respond with fitting messages. For example, an AI tutor might notice frustration or confusion in a student's questions. It could then offer encouraging feedback or ask a clarifying question in a gentle tone. The AI does not feel empathy, but it can produce empathetic responses that suit the situation. In some respects, it can be more consistent than a human. An AI does not become impatient or burned out. It can offer support at any hour, which helps students who need steady reassurance or a listener who will not judge them. AI systems can also track a student's emotional patterns over time. They may flag a decline in engagement or signs of distress that a busy teacher could miss. Such long-term monitoring could allow earlier help for students who are struggling.

Studies from 2024 and 2025 support AI's competence in emotional support. A review of AI and well-being in higher education describes a study of 401 Chinese university students. These students believed that AI could improve mental well-being. The researchers note that AI chatbots and virtual assistants give immediate support and essential information, which makes mental health services available to more people [7]. The first randomized controlled trial of a therapy chatbot based on generative AI showed clear clinical benefits. Compared with controls, participants showed greater reductions in symptoms of major depressive disorder (mean change −6.13 vs. −2.63 at 4 weeks), generalized anxiety disorder (−2.32 vs. −0.13), and high risk for feeding and eating disorders (−9.83 vs. −1.66) [8]. Participants used the chatbot, Therabot, for more than 6 hours on average. They rated their therapeutic alliance with it as comparable to that with human therapists [8]. A qualitative study interviewed nineteen people who used generative AI chatbots for mental health. They reported high engagement and positive effects, including better relationships and healing from trauma and loss. Four themes emerged from the interviews: a sense of "emotional sanctuary," "insightful guidance" (especially about relationships), the "joy of connection," and comparisons between the "AI therapist" and human therapy [9].

People, including students, are starting to use AI for emotional support on demand. Anthropic reports that most use of its model Claude is task-oriented. About 2.9% of conversations, however, are "affective conversations," in which users seek advice, counseling, or companionship [10]. In these conversations, people often discuss personal challenges such as relationships, loneliness, and anxiety, and they receive advice or coaching. The researchers also found that users' expressed sentiment tends to become more positive over the course of a coaching or counseling conversation. This suggests that the AI's responses often improve the user's mood during the exchange [10]. Podcast discussions offer similar anecdotes. Guests on *The Ezra Klein Show*, for instance, have described teenagers who talk to AI "friend" apps late at night when they feel they have no one else [11]. Such accounts show both the opportunity and the concern in this domain. An AI companion that is always available can reduce loneliness. Heavy reliance on AI for companionship, however, could create new forms of social isolation or confusion about what is real [10]. Mollick imagines a future in which "AI companions become far more compelling to speak with than most other people, and ... loneliness becomes less of an issue." In the same future, "some people would rather interact with AIs than with humans," which raises difficult questions for society [2].

### *2.1 What makes AI emotionally competent: technical mechanisms*

Benchmark studies measure this competence directly. Six large language models (ChatGPT-4, ChatGPT-o1, Gemini 1.5 Flash, Copilot 365, Claude 3.5 Haiku, and DeepSeek V3) outperformed humans on five standard emotional intelligence tests. The models reached an average accuracy of 81%. The human average in the original validation studies was 56% [12]. A separate study found that GPT-4 can identify emotions and reason about managing them. The same study found that GPT-4 lacks deep reflective analysis of emotional experience and of the motivational side of emotions [13]. AI does not experience emotions, but it can recognize them and respond to them effectively.

AI's emotional competence rests on the transformer architecture and its attention mechanism. Self-attention lets a model look at a whole input sequence at once and decide which parts deserve focus [14]. In a conversation, this allows the AI to find emotional cues in different turns and to see how they relate. Self-attention also captures long-range links within and across modalities such as text, audio, and images [15]. This matters for emotional context, which often spreads over several exchanges. In emotion recognition, attention lets the model focus on the most relevant parts of each input. The model weighs different features of the data, so it can capture the subtle emotional signals that accurate recognition requires [15].

A training method called Reinforcement Learning from Human Feedback (RLHF) also plays a central role. RLHF adds human feedback to the reward function, so the model learns to act in ways that better match human goals and needs [16]. In emotional support, human raters can score responses on human qualities such as friendliness, a fitting degree of context, and mood [16]. One analysis of ChatGPT links its human-like responses to this combination of the transformer architecture and RLHF [17]. Through this training, AI learns subtle patterns of emotional interaction that would be hard to program by hand. A systematic review describes the result as a form of cognitive empathy. The models recognize emotions and give emotionally supportive responses across many contexts [18].

The optimistic view is that AI can add to emotional support in education, while human care remains central. AI can give individual encouragement, simulate empathy, and offer coaching. An AI learning app, for example, might encourage a struggling student ("I know this is tough, but I believe you can do it!"). It might suggest a short break when it senses frustration. It might also help a shy student practice social skills by role-playing difficult conversations in a safe setting. Some counseling centers have tested AI chatbots with college students for cognitive-behavioral exercises and mood tracking. Such tools must be designed and monitored with care. Ethical guidelines are needed so that the AI stays supportive and avoids harmful advice. Anthropic's safety work aims to align models with human values. For instance, Claude is designed to avoid exploiting users' emotions and to avoid encouraging delusions [10]. In practice, the best results may come from human-AI collaboration in pastoral care. Teachers and counselors could use AI to find students who need help and to extend support to more of them. Humans would provide the authentic empathy and judgment that machines lack. With this balance, AI's constant presence and pattern recognition could strengthen the emotional safety net for students. Fewer students would go unnoticed.

## 3. AI and Creativity

Another common belief holds that AI is good at repetitive or analytical work and cannot foster human creativity. Creativity in schools has usually grown under inspiring teachers. These teachers encourage divergent thinking, art, play, and intellectual risk. A software program may seem an unlikely aid to these processes. Evidence suggests, however, that AI can strongly support creativity. It can produce new ideas itself, and it can stimulate creativity in students. Modern AI models are trained on huge datasets that cover literature, art, history, science, and more. They can process and recombine this information in new ways. Mollick calls them "connection machines" because they link ideas across domains in unexpected ways [2]. A person may be limited by personal experience or by habit. An AI can offer an unusual analogy, an odd suggestion, or a hybrid concept that sparks a learner's imagination. For example, a student brainstorming a story might ask an AI for ideas. The AI could propose a plot that mixes ancient mythology with futuristic science, a combination the student had not considered. Such surprises can challenge norms and spark human creativity. A well-designed AI can form genuinely new combinations and concepts from the knowledge it has absorbed [19]. AI can also generate and evaluate many scenarios quickly, which promotes divergent thinking in students [20]. When the AI proposes many variations or solutions, students move past their first idea. They consider alternatives and widen their creative range [20].

### *3.1 Recent evidence of AI's creative competency*

Research from 2024 and 2025 shows AI's creative competence in several domains. One large study analyzed more than 4 million artworks by more than 50,000 users. It found that text-to-image AI raised human creative productivity by 25%. It also raised the value of the work by 50%, measured as the likelihood of receiving a "favorite" per view [21]. Peak content novelty increased over time. The artists who gained the most were those who explored new ideas and filtered model outputs for coherence. Human ideation and artistic filtering therefore decide how much an artist gains from generative AI tools [21].

Studies in education show similar results. In a 2024 study of college students, every participant found AI helpful for brainstorming. Students produced more diverse and more detailed ideas when they used AI [20]. AI helped start creative sessions. It also acted as a partner that did not judge, so students explored ideas they might hold back in a group [20]. A 2025 industry report found that 91% of educators see better learning when their students use creative AI tools [22].

AI can also match or exceed human performance on standardized creativity tasks. One study tested ChatGPT-4 on the Figural Interpretation Quest (FIQ), a task that asks for creative interpretations of images. On average, the AI showed greater flexibility than humans in its interpretations. The evaluation also revealed subtle differences in how creativity appears in human and artificial systems [23]. In a controlled experiment on short-story writing, access to ideas from generative AI made stories more creative, better written, and more enjoyable, as judged by evaluators. The effect was strongest among less creative writers [24]. The same study found that AI-assisted stories were more similar to one another. This reduces the collective diversity of new content [24].

Other test results point the same way. The Alternative Uses Test asks people to list as many uses as they can for a common object. Advanced AI systems score very well on it. Mollick reports that GPT-4 outperformed more than 90% of human test-takers in generating creative uses for everyday items, as judged by human evaluators [2]. Such tests have limits, since the AI may have seen examples during training. They still show that AI can produce a range of imaginative ideas comparable to that of very creative people [2]. AI creativity also extends beyond text. Generative models for images and music, such as DALL-E, Midjourney, and MuseNet, produce original images and compositions. Teachers on podcasts have described students who use AI art generators to visualize their ideas. The images then inspire the students to refine or reimagine their projects. These tools often motivate students to iterate and experiment more boldly. Guests on the *Latent Space* podcast, for example, have noted that generative AI lets novice programmers and designers prototype creative projects, such as games or animations, quickly. This lowers the barrier to creative expression. Kevin Kelly describes this widening of creativity, in which every student has a tireless creative assistant: "It's not a race against the machines. You'll be paid in the future based on how well you work with robots… Robots will do jobs we have been doing, and do them much better… And they will help us discover new jobs… They will let us focus on becoming more human than we were" [3]. In the classroom, "becoming more human" means focusing on the human parts of creativity. Students define the problems, bring empathy and values to the work, and make aesthetic and ethical choices. The AI handles tedious details or generates raw material for inspiration.

## *3.2 Technical mechanisms enabling AI Creativity*

The computational basis of AI creativity also begins with the transformer architecture. Its self-attention mechanism processes input text and finds relationships between elements that lie far apart in the data [14]. This supports what researchers call combinatorial creativity. In this process, AI systems find, retrieve, and recombine relevant concepts from several domains in a structured way [25].

Computational creativity research describes creativity as "a search process through the space of possible combinations" [26]. The combinations can come from joining different representations. They can also come from rule-based or random transformations of starting and intermediate representations [26]. Large language models perform this search through two main features. First, their stochastic output and the variety of prompts they receive often lead to novel results [19]. Second, their training on very large datasets lets them reach and recombine elements from the whole range of knowledge they have seen [19].

Recent studies show that LLMs can carry out combinatorial creativity in a structured way [27]. In one study, LLMs explored ideas from the scientific literature. They analyzed conceptual relationships across papers and found new research opportunities [27]. The process builds connections across domains. Ideas are embedded and compared at several levels of abstraction, so the system can join ideas from unrelated fields. Structured formats record how each retrieved idea relates to the abstract problem, which keeps the process traceable [25].

Attention helps creativity in a second way. A model can weigh every part of a long text when it decides what to generate next [14]. It can therefore detect patterns and relationships that span long distances. Unexpected links of this kind are typical of creative thinking. Multi-head attention adds another layer. Several attention heads work in parallel, and each can track a different type of relationship in the input. Together they capture the varied patterns that feed creative output [14].

Training methods also shape AI's creative abilities. RLHF fine-tunes models to match human preferences and values, including creative preferences [28]. The model learns what is statistically likely, and it also learns what people find creative, valuable, and engaging. Other techniques, such as diffusion models and generative adversarial networks, produce new content that stays coherent while it explores new possibilities [28].

Classroom use of AI for creativity has already begun. Some educators use AI as a brainstorming partner for students. In a group project, for example, a language model might play the role of a "wild idea generator." One teacher described a design challenge in which students critiqued and built on AI-generated solutions. The unusual AI suggestions pushed students to think in new directions. The students' final designs were more innovative than those of earlier semesters. Mollick has made AI use mandatory in some entrepreneurship assignments. He asks students to attempt projects far beyond their normal ability, because AI is available to help [2]. One of his prompts reads: "Make what you are planning on doing ambitious to the point of impossible; you are going to be using AI… I won't penalize you for failing if you are too ambitious" [2]. As a result, students aim higher. They may try to build a working app or write a business plan in a week. They often achieve more than they expected, and they learn a great deal about creative problem-solving. AI provides the scaffolding: it writes code snippets, drafts marketing copy, and simulates user feedback. Students remain the directors of their projects. They exercise higher-level creativity when they select, refine, and combine the AI's contributions. In this sense, AI acts as a creative muse or apprentice. It widens the range of ideas and handles low-level execution. Students are then free to focus on imaginative and integrative thinking [21][24].

Using AI for creativity in education requires careful guidance. Teachers must make sure that students evaluate AI outputs and avoid plagiarism. Mollick points to a paradox in AI creativity. The same randomness that lets AI produce new ideas can also produce misinformation or unfiltered content [2]. Educators therefore have a key role. They coach students to judge AI-generated content critically and to improve it. Used well, AI can greatly amplify creativity in the classroom. It widens what students can imagine and make. It also prepares them for a future in which creative work with AI is normal.

## 4. AI and Contextual Understanding

Teachers are often praised for understanding each student's personal, social, and cultural context. Many factors outside test scores affect learning. A common view holds that AI lacks this holistic understanding. On this view, AI cannot truly personalize learning or respond to individual needs as a human teacher can. Human educators do use intuition and experience to read subtle cues. They notice a student's body language, family and community background, and current events in the student's life. Advanced AI systems, however, can increasingly analyze large and varied streams of data to build a detailed picture of a learner's context. AI's contextual understanding comes from its ability to process and combine many data streams at once. In principle, an AI tutor could draw on a student's academic records, past homework, data from educational games, and even biometric or environmental data, where such data are available and ethically obtained. It could use these inputs to assess the student's state and needs continuously. Consider an AI learning platform that records which topics a student finds hard and how quickly the student answers at different times of day. It might also note which concepts excite the student, based on engagement patterns. It might even receive data from a fitness tracker about sleep or stress. By combining these inputs, the AI can build a more detailed and current model of the student than a teacher with 30 students has time to build. AI can analyze academic performance, social interaction, physiological data, and environmental factors together. This allows careful interpretations of behavior and performance that human observers might miss. For instance, an AI might find that a student does better on math exercises in the morning and struggles after 2 p.m. If it also knows that the student has sports practice in the afternoon, it might suggest doing math homework earlier in the day. It might also detect that a normally active student has stopped asking questions in online discussions, and flag a possible loss of engagement.

### *4.1 Recent evidence of AI's contextual competency*

Research from 2024 and 2025 shows that AI systems have become highly competent at contextual understanding. They achieve this through data integration and adaptation. A bibliometric analysis of adaptive learning technologies finds that AI systems use data analytics and machine learning to personalize instruction. They adapt content, pace, and delivery to each student's strengths, weaknesses, and preferences [29]. Adaptive learning platforms lead to higher pass rates and better student retention than traditional teaching methods [29]. These results show that AI's contextual understanding works in practice.

Multimodal AI systems mark a major advance in contextual understanding. Large multimodal foundation models can process speech, music, images, and video at the same time [30]. They can therefore build a full contextual picture from varied data. Their contextual awareness also makes cognitive tutors more effective with ill-defined problems [30]. Google Gemini is one example of a multimodal generative AI tool. It processes text, image, audio, and video inputs and produces many types of content [31]. It can create differentiated materials and design activities for students at different levels. It can also give extra explanations to students who need more support. At the same time, it analyzes learners' work to give personalized feedback and to identify areas for improvement [31].

A 2025 study in STEM learning environments shows how multimodal analytics can support collaboration among teachers, researchers, and AI. The AI produced multimodal timelines that combined many data types [32]. These included students' emotional responses, synergy scores, measures of social interaction, prosodic audio cues, verbatim transcripts, prior knowledge of physics and computing, and detailed learning analytics. The combined view gave teachers new insight. It helped them identify student difficulties and write supportive feedback [32].

Adaptive learning will likely use even more contextual information in the future. Systems may draw on wearable devices, environmental sensors, and other sources. They could then adapt content to factors such as location, time, or the learner's emotional state [33]. Context-aware adaptation lets e-learning systems offer content that is more relevant to each learner and helps learners complete their tasks [33]. These systems also increasingly include collaborative and social learning. Their algorithms analyze learner interactions, group dynamics, and social network data. They then recommend group projects, collaborative activities, and peer feedback for each learner [33].

Anthropic's Education Report shows AI's contextual skill in real use. The study analyzed more than half a million student conversations with Claude. It found distinct patterns in how students work with AI. One pattern was "collaborative problem solving," in which student and AI go back and forth until the student understands [5]. Students used Claude across many disciplines and tasks, from debugging code to explaining legal concepts. They often asked for explanations suited to their own level of prior knowledge [5]. Students mainly used AI to create and improve content, and these tasks required an understanding of context. Examples include drafting essays with personalized feedback and getting study help for specific courses [5]. Claude can sustain long conversations and refer back to earlier parts of the dialogue. In effect, it remembers what the student said or found difficult before. AI tutors can therefore simulate the continuity of a human tutor who follows a student's progress over weeks and months. Anthropic's Claude for Education initiative stresses this ability. It offers institutions a secure AI that can bring documents, tools, and web knowledge into a conversation [34]. A student can connect Claude to project documents or class notes. Claude then uses that material when it answers the student's questions. Its answers therefore fit the student's current curriculum and situation. This ability to connect context is a major step for personalization. The AI gives answers that reflect what the student is working on, and it can adapt as the work changes.

## *4.2 Technical mechanisms enabling contextual understanding*

The architecture of AI explains much of this contextual understanding. The transformer and its attention mechanism were introduced in 2017 in the paper "Attention Is All You Need" [35]. Attention is a machine learning technique that directs a model to the most relevant parts of its input. It builds sensitivity to context directly into the way the model represents information [36].

Self-attention lets the model consider every part of the input when it generates a response, whatever the position of each part [37]. The mechanism calculates relationships and dependencies between tokens, even those far apart in the input. The model can therefore understand how words at the start of a context relate to words at the end [37]. Self-attention computes weights that show how relevant each token is to every other token. Each element of the input attends to all the others, so the model captures global dependencies [38].

In practice, contextual understanding depends on the context window. The context window is the amount of information an AI system can consider at once. Modern systems have greatly expanded their context windows. At the time of writing, Gemini 1.5 could process up to 1 million tokens, and Anthropic's Claude models offered windows of 200,000 to 500,000 tokens [39]. Such large windows let AI systems keep track of long conversations, long documents, and other context at the same time.

Multi-head attention strengthens contextual understanding further. The model runs several self-attention operations in parallel. Each one has its own learned query, key, and value transformations [36]. The model can therefore capture different kinds of relationships between words at the same time. Each attention head learns its own projections and may focus on a different type of relationship, such as syntactic, semantic, temporal, or contextual links [36].

Multimodal systems apply the same mechanisms to several kinds of input. Their transformer-based architectures align and process text, audio, and visual data together [40]. Such a system can process students' emotional responses, audio cues, conversation transcripts, and visual behavior at the same time. It can then build a full model of each student's learning state [32].

Contextual awareness also depends on how a system manages memory and state. Older sequence models processed information one step at a time. Transformers can attend to all parts of the input together [36]. This supports what researchers call "contextual continuity." The system can keep and use contextual information across long interactions, much as a human tutor remembers a student's progress over time [40].

Kissinger and colleagues argue that AI systems may become "ultimate polymaths." Such systems would draw knowledge from many domains at once as they explore the frontiers of human understanding [1]. In education, this quality lets an AI tutor bring history, culture, or a student's local environment into a lesson as it unfolds. Imagine a class discussion of a novel. A context-aware AI could supply background on the novel's historical setting, adapted to the student's region. It could also link the novel to current events that matter to students of that age. Even well-read teachers find it hard to do this in real time for every student. Mollick adds that AI works more like a person than like a rigid program. It can handle nuance and ambiguity in conversation, and both are key to understanding context. One of his principles of "co-intelligence" is to "always invite the AI to the table," as one would invite a collaborator who can contribute ideas [2]. On this view, AI is a partner that keeps taking in context, such as the ongoing discussion and the user's goals, in order to add value.

Early classroom trials show promising results. At some universities, instructors report that AI writing assistants give students highly individual feedback. The assistants adjust their suggestions to the assignment and to the student's earlier drafts. Consider a student who finds it hard to link biology concepts to real-world examples. An AI tutor that remembers this from earlier sessions could remind the student: "Recall how we compared cell structure to a factory last time; let's try a similar analogy here." Such capabilities point to personalized learning at a scale that was previously out of reach. Nick Joseph, Anthropic's head of training, has said that AI could give every learner an individual learning experience, much like a personal tutor, and that this would bring major changes to education [41]. He suggests that by the time today's young children reach school, each of them may routinely have an AI tutor that adapts in real time to their context and needs [41].

Several challenges remain. AI's contextual understanding is only as good as its data and the patterns it can detect. Algorithmic bias is one risk. If the data about a student's context are incomplete or biased, the AI's recommendations may miss the mark or even deepen inequities. Privacy is another concern, since extensive personal data in AI systems raise ethical issues. Responsible use of context in AI tutoring requires strict data governance. It also requires transparency, so that students and parents know how the AI "learns" about them. If these challenges can be managed, the potential benefit is very large. Instruction could be truly differentiated, with content, pacing, and strategy adjusted continually to each student's context. Educational psychologists have long held this aim. Bloom showed that one-to-one tutoring can raise student achievement by about two standard deviations, the "2 sigma" effect [42]. AI may finally bring every student close to that ideal. It combines a wide view of context with precise personalization in the moment. AI can thus reach a form of holistic understanding. This understanding differs from human intuition, but it is powerful in its own systematic way, and it lets AI support each learner individually.

## 5. AI and Student Engagement

Keeping students engaged and curious is a lasting challenge in education. Great teachers use charisma, storytelling, interactive activities, and personal rapport to "ignite a love of learning." Critics claim that AI tutors or AI-based curricula can never do the same. In the popular view, AI-based instruction would be sterile and dull, built on dry drills and isolated screen time, and it would fail to inspire. Evidence from adaptive learning systems and AI-driven educational games points the other way. AI can spark curiosity and sustain engagement by fitting content and teaching methods to each learner. Recent research supports this view. Studies of AI in higher education report clear benefits, including personalized learning, better outcomes, and higher engagement. Adaptive platforms raise engagement and performance through real-time feedback and customized learning paths [43][44]. Industry data from 2024 indicate that 54% of students engage more with their coursework when AI tools are part of the learning experience [45].

The main advantage of AI here is fine-grained personalization. AI can keep analyzing a student's responses, learning patterns, and interests. It can then adjust the material to keep the student in an optimal state of challenge and interest. This matches two established ideas in educational psychology. Vygotsky's zone of proximal development (ZPD) holds that students learn best on tasks just beyond their current ability, with suitable support. Csikszentmihalyi's flow theory holds that deep engagement occurs when challenge and skill are well matched. Studies from 2024 apply these ideas to AI systems. One finds that "AI tools assist the students in identifying and operating within their ZPD" [46]. Another finds that AI applied within the ZPD can give students personalized resources and help, and can offer teachers content and strategies suited to their students' needs [47]. AI tutors are well placed to keep students in this zone. They can detect at once when a task is too easy, as when a student races through without errors. They can also detect when it is too hard, as when errors pile up or the student pauses for long periods. They then adjust the difficulty in real time. Machine learning algorithms make this possible. They assess each student's current knowledge, pace, and preferences, and they adjust content and difficulty so that the work is neither too easy nor too hard [48]. An AI math tutor, for example, might give quicker and more subtle hints when a student struggles. This prevents frustration and keeps the student moving. If a student finds the material trivial, the AI can skip ahead or add a harder extension problem to revive interest. A human teacher cannot adapt in this way for 30 children at once. AI can manage it for each student through constant monitoring and feedback.

AI systems also borrow techniques from game design and behavioral psychology to sustain engagement. These include points, badges, immediate feedback, and elements of surprise. AI-driven platforms can build reward structures and feedback loops that encourage students to continue. Research on AI-enhanced gamification finds that AI tools can change how students learn through personalized feedback, adaptive pacing, and targeted learning experiences [49]. A systematic review adds that gamification improves motivation, engagement, and skill acquisition. The same review finds that AI personalizes gamified content and adapts it to each learner [50]. An AI tutor might celebrate milestones ("Great job on 5 in a row!") or set small challenges ("Try to beat your previous time on this puzzle"). A well-designed game keeps players involved in the same way. AI tutors can also present content in unexpected ways or draw new links between topics. Surprise and discovery are key to curiosity. A literature tutor might link a theme from Shakespeare to a current film the student enjoys. The student may think, "Wow, I never thought of that!" and feel a fresh burst of interest. AI can also learn what sparks each student's curiosity and use it across subjects. Adaptive systems choose a format for each learner. Visual learners might receive infographics, while hands-on learners might receive interactive simulations. Shared goals and collaborative challenges can also raise engagement among peers [51]. AI may notice, for instance, that a student responds well to real-world applications. It can then frame abstract math problems in terms of sports or shopping if those topics engage the student. Such personalization can reveal hidden interests or talents. It shows students connections they might never have met otherwise.

Real-world observations are beginning to confirm these advantages. Soon after ChatGPT was released, Ethan Mollick noticed a change in his university classes. Students raised their hands less often to ask factual or definition questions, because they could get those answers instantly from AI [2]. This might look like lower engagement. Mollick reads it positively. AI handled routine questions, which freed class time for deeper discussion and problem-solving [2]. Students came to class with a higher baseline understanding and more curiosity about complex issues. Mollick changed his teaching in response. He now expects students to use AI for a first step, such as a draft or initial research. Class time then goes to higher-order critique, debate, and application. His in-person sessions have become more engaging as a result, because teacher and students can explore more interesting ground. Hosts of the *Hard Fork* podcast have also noted that many universities are moving from banning AI to using it productively. Their aim is to make classes more engaging and relevant in a world full of AI. Khan Academy's early trials of its AI tutor, Khanmigo, offer an example from K-12 education. Sal Khan reported that students who used Khanmigo spent more time working through ideas, because the AI kept prompting them with questions. Teachers also found that these students were often more eager to share what they had learned. This points to higher engagement and stronger ownership of learning.

Controlled studies now support these reports. A 2024 study of AI-driven personalized learning and intelligent tutoring systems involved 300 students. It found "significant improvements in both engagement and academic performance." Mean engagement scores rose from 3.5 to 4.2 ($p < 0.001$) [52]. A systematic analysis of studies from 2007 to 2024 found that AI works well as a tutor. It gives personalized support and adaptive feedback that guide students through problem-solving and creative learning [53]. The same analysis reports that AI platforms improved students' self-regulation and knowledge building. They gave real-time answers to students' questions and reduced interruptions, which helped students stay emotionally engaged [53].

Anthropic's Learning Mode for Claude aims directly at engagement and deep thinking. In Learning Mode, Claude guides students with questions such as "How would you approach this problem?" It also uses Socratic questions such as "What evidence supports your conclusion?" These prompts lead the student to think and explain [34]. The mode puts guidance ahead of answers, so the student stays mentally active throughout the exchange. Master teachers work in the same way. They answer a question with another question that helps the student find the answer. Students who build knowledge in this active way tend to engage more and remember more. Learning Mode also stresses core concepts and offers templates and study guides that help students organize their thoughts [34]. A study of undergraduates in China supports this approach. Students generally valued AI-generated feedback, especially when it was specific, clear, and corrective. They also believed that AI could analyze large amounts of data to create personalized learning paths and so improve their learning [54]. These features match known drivers of engagement: clear goals, suitable challenge, interactivity, and immediate feedback.

Several computational approaches underlie AI's effect on engagement. Machine learning algorithms optimize learning paths. Decision trees choose the best next activity based on performance. Neural networks find complex patterns in learning behavior. Clustering algorithms group students with similar learning styles [33]. Attention mechanisms let AI systems learn how much each past question matters for predicting a student's current response. They also model forgetting by taking into account the timing and frequency of past practice [55]. Interactive content, such as videos with embedded quizzes and game-like elements, sustains learners' interest and motivation. Affective computing, sentiment analysis, and facial expression recognition let systems detect students' emotional states during learning. The system or the teacher can then respond with timely help and personalized feedback [56].

AI can also help teachers make their own lessons more engaging. AI tools let instructors create interactive materials, such as case studies, simulations, and fresh examples, whenever they need them. College teachers have used GPT-4 to invent relatable metaphors or humorous analogies that open dry topics and capture attention. Mollick observes: "AI is very good at assisting instructors to prepare more engaging, organized lectures and make the traditional passive lecture far more active" [2]. Institutions are moving in the same direction. A 2025 survey of more than 800 higher education institutions found that 57% were prioritizing AI in 2025, up from 49% the year before. These institutions are investing in technology, data-driven strategies, and digital learning environments to make education more adaptable and individual [57]. AI takes over part of the preparation and supplies creative ideas. Teachers can then focus on leading lively discussions and hands-on activities, which engage students naturally.

Further evidence on engagement continues to appear. Industry statistics suggest that personalized learning with AI can raise retention rates by as much as 30% [45]. In one educator survey, 25% of respondents named help with personalized learning as a benefit of AI, and 18% named improved student engagement [58]. A 2024 summary of research on personalized learning concludes that it is worth the investment, because students learn more and stay interested [59].

Thoughtful use of AI can make it a strong driver of student engagement. Personalization, real-time adaptation, game design, and Socratic guidance keep learners in the best zone for curiosity and concentration. A single lecture for everyone leaves some students bored and others lost. With AI, each student can have a tailored experience that holds their interest. The human educator's role shifts toward orchestrating these individual paths. AI handles routine interactions and frees time for the most engaging human-led experiences. Current practice points to an optimistic picture. Every student is deeply engaged, either with an AI tutor or in group work. The teacher moves around the room and offers insight, encouragement, and a human touch that no machine can replace.

## 6. AI and Problem-Solving Skills

Critical thinking and problem-solving have long been hallmarks of a good education. A skeptic of educational AI might say: "Sure, an AI can give students answers or solve problems for them, but it won't teach them to solve problems themselves." Many fear that reliance on AI will turn students into passive consumers of solutions and weaken their own problem-solving. Research confirms that some of these concerns are real. Studies link "increased reliance on artificial intelligence (AI) tools" to "diminished critical thinking abilities" [60]. Younger participants, aged 17 to 25, depended more on AI tools and scored lower on critical thinking than older groups [61]. A broader analysis gives a more balanced picture. AI "presents challenges, such as over-reliance on technology, diminished critical thinking, and the risk of academic fraud." It also "offers significant benefits, such as personalized learning, improved educational outcomes, and increased student engagement" [62]. Proper use of AI can strengthen problem-solving. A well-designed AI learning system models and scaffolds the problem-solving process. It coaches students in critical thinking one step at a time.

AI develops problem-solving skills by modeling and analyzing complex problem spaces with precision and depth. A human teacher may prefer one way of approaching a problem. An AI can generate and evaluate several solution paths at once. Research on the metacognitive demands of generative AI supports this view. AI systems can show users "different and perhaps surprising perspectives." They can also support "more flexible and self-aware problem-solving," for example when users choose a temperature setting that balances diversity and factual accuracy [63]. An AI tutor can therefore show students many problem-solving strategies in a short time. A single teacher or textbook often cannot do this. Given a physics problem, for example, an AI might solve it by an algebraic method, a graphical method, and a simulation. The student can compare the methods and understand each one. By seeing different strategies, students learn that many problems can be approached from several angles. This is a key part of creative problem-solving.

AI can create an almost endless variety of practice problems suited to each student's current level. Students are therefore challenged without being overwhelmed. This adaptive generation of problems is vital for building skill. Once a student masters a concept, the AI can present a slightly harder problem or a new twist. It can also revisit earlier material in new contexts to strengthen transfer. Machine learning makes this possible. Adaptive systems gather, analyze, and interpret large amounts of learner data. They detect patterns, identify strengths and weaknesses, and generate personalized recommendations and interventions [33]. Newer systems can also generate plans automatically from text descriptions of students' actions in a game-based learning environment. These plans provide adaptive scaffolding for goal setting and planning, which are central to self-regulated learning [64]. Adaptive levels and immediate, detailed feedback on each step together form a strong framework for developing problem-solving skills.

AI tutors are especially good at instant feedback, which traditional homework rarely provides. With traditional homework, a student often struggles alone and receives corrections days later. An AI can point out a mistake at once ("Check your calculation at step 3, it seems off") and ask the student to reconsider. It can also give a hint when the student is stuck ("Have you tried drawing a diagram of the problem?"). Research supports this approach. Timely feedback lets students adjust their understanding and approach [65]. AI systems can assess student work instantly and give immediate feedback, which allows quick correction and a faster learning cycle [66]. Scaffolding adds further benefit. In one study, AI agents that asked proactive scaffolding questions produced significant gains in student comprehension, and the gains lasted beyond the intervention [67]. Such interactive guidance resembles one-to-one tutoring, which research has shown to be one of the most effective ways to build problem-solving skill.

Mollick (2024) shows how AI can help students take on ambitious projects that combine problem-solving across domains. In his assignments, students use AI tools to finish in days work that would normally take weeks or months. Examples include building a working app prototype or analyzing a complex dataset. Students often achieve more than they thought possible, because AI handles some of the drudgery and supplies expertise on demand [2]. In these projects, students direct the AI, evaluate its outputs, and improve their solutions step by step. These are themselves higher-order problem-solving skills [2]. Empirical research supports these observations. A 2024 study of 300 high school students who used intelligent tutoring systems found significant gains in problem-solving (pre-test $M = 65.4$, post-test $M = 72.8$, $t(299) = 4.67$, $p < 0.001$). It also found gains in critical thinking ($M = 68.9$ to $74.3$, $t(299) = 3.82$, $p < 0.001$) and in logical reasoning ($M = 63.2$ to $70.1$, $t(299) = 3.45$, $p = 0.001$) [52]. Consider a student who uses AI to generate code that then fails to run. The student must troubleshoot by reading error messages, with help from the AI, and by adjusting the approach. The AI serves as a cognitive apprentice. It helps with lower-level tasks and with finding information. The student can then focus on learning how to solve the problem.

With repeated guided practice, students internalize these patterns. AI can identify patterns in a student's approach and highlight strengths and areas for improvement. It can help the student develop a systematic method that transfers to new situations. Educational research supports this process. AI systems can scaffold learning to strengthen critical thinking. They present tasks within the student's zone of proximal development, and they offer personalized experiences that adapt to each student's style and ability [68]. For example, the AI might notice that a student tends to skip planning and jump straight into calculations, which leads to errors on complex problems. It could then ask the student to outline the steps first. It could also show how breaking a problem into smaller parts leads to more success. In this way, the student builds better habits.

Large language models also hold what researchers call "metacognitive knowledge about mathematical problem-solving." AI systems can be designed to extract and use the models' implicit knowledge of mathematical skills and concepts, which improves their problem-solving [69]. Studies of AI-based scaffolding find that adaptive learning technologies support cognitive and emotional engagement between students and course content [70]. In these systems, personalized feedback builds on personalized scaffolding. It uses each student's strengths and weaknesses to give targeted feedback at once [71].

Anthropic's Claude, with its step-by-step reasoning, is designed to give clear and structured help with complex questions [34]. When a student asks Claude a multi-step question, it often breaks the solution into logical steps ("First, let's define the problem… Next, consider this factor… Now we do this calculation…"). By modeling this structure, the AI teaches the student how to think through a problem. A student may eventually internalize that voice and ask, "What would Claude ask me to consider next?" The AI then becomes a metacognitive coach that teaches self-questioning. Research on AI literacy courses supports this view. Courses that stressed problem-solving competence led to "a significant improvement in their metacognitive strategies in problem-solving." Students also gained a better understanding of the ethical limits and principles that govern the use of AI for problem-solving [72]. Anthropic's Learning Mode, described above, deliberately guides students toward answers, which is vital for developing problem-solving skill [34]. In Learning Mode, the AI might say, "How might you break this problem down? Let's try tackling one part at a time." Such prompts lead the student into the actual work of solving the problem. Over time, students learn to ask these questions themselves. Independent problem-solving is the final goal.

Advanced AI systems now use explicit teaching methods. One example is the "Socratic Playground for Learning," which uses the Socratic method to foster critical thinking [73][74]. It creates specific learning scenarios and supports multi-turn tutoring. It also provides adaptive scaffolding by combining symbolic knowledge representations with neural learning. This lets it address learner misconceptions precisely and support step-by-step cognitive development [73][74]. In computer science education, AI can support metacognitive skills and reflective learning. Students explain their reasoning to the AI in detail and organize their solution strategies. The AI offers a different perspective on the problem, which stimulates reflection [75].

Podcasts such as *The Cognitive Revolution* have featured education innovators who report early success with AI tutoring. One example was an "AlphaCode Club," in which middle school students use AI coding assistants to solve programming challenges. Students who use the AI solve more challenges. They also learn to debug and refine code through their exchanges with the assistant, which is a valuable skill in computer science. When a program fails, they now test and fix problems systematically, because the AI has prompted them to do so. Teachers had struggled to build this persistence and method before. AI can therefore increase student persistence. It gives timely hints and moral support ("No, that didn't work – but don't worry, debugging is normal. Let's print out this value and see what's happening.").

Experimental research supports these accounts. A quasi-experimental study of 120 engineering students tested ChatGPT-assisted instruction based on a Constructivist Inquiry-Based Learning Prompting (CILP) framework. Students showed significant gains in conceptual understanding [76]. The AI gave "dynamic and adaptive scaffolding through real-time, dialogic interactions." This helped students develop metacognitive skills for overcoming misconceptions [76]. Research in programming education points to one technique as the most effective. The learner works through the problem step by step in a dialogue with the AI and states what needs to be done at each stage before the code is revealed. This approach best helped students apply concepts later without AI assistance [77].

Balance remains essential. Teaching metacognitive skills helps students judge the quality and reliability of AI outputs. Some assignments should require problem-solving without AI, so that students practice independent thinking and AI supports human reasoning without replacing it [78]. Without proper guidance, students can become over-reliant on AI. They may accept solutions blindly or use AI as a crutch without reflection. Educators are developing strategies to prevent this. Some ask students to "critique the AI's solution" as part of an assignment. Others give the AI slightly wrong inputs on purpose, so that students practice checking and correcting [2]. Researchers also suggest creating educational content that may contain errors and asking students to act as tutors who correct the AI's mistakes. This builds appropriate reliance on AI-generated content [79]. Mollick uses a similar exercise. Students generate an essay with AI and then must identify its weaknesses and errors, which forces them to engage critically with the output [2]. Another strategy is closed-book exams or in-class work without AI. These ensure that students have mastered the underlying skills on their own. Assessment may therefore move toward such settings, while AI remains a tool for learning and practice.

Real-world implementations support these teaching approaches. Educational platforms report that AI tutors adapt to each student's learning style and pace. They provide customized lessons and feedback that help students understand and remember concepts [80]. Systems such as Khanmigo show that effective AI tutors guide learners to find the answer themselves [81]. Technical analyses describe the components of modern AI tutoring systems. Natural language processing supports smooth communication. Assessment and feedback modules monitor progress. Adaptive learning technology customizes lessons based on each student's performance [82][83].

Thoughtful integration makes AI a strong tutor for problem-solving. It offers unlimited practice at adaptive difficulty, immediate and detailed feedback, many solution strategies, and a patient dialogue that models how to think through problems. Human-AI collaboration in complex problem-solving has been studied across many application domains. AI systems can support its cognitive, metacognitive, social, and affective dimensions [84]. These are the conditions under which students' own problem-solving abilities grow. Early evidence suggests that students who learn with AI support can become more independent problem-solvers. Scaffolding has led them to success, and they have seen what effective problem-solving looks like. They gain confidence and a set of approaches they can use. On the optimistic view, AI speeds the growth of analytical and critical thinking. It acts as a personal mentor that constantly challenges and supports students. When students move on to real-world problems, they bring these skills with them. They may also keep using AI as a partner in professional problem-solving, as many engineers and scientists are beginning to do. Education can then produce graduates who use AI well and who are also strong problem-solvers in their own right. Such graduates can check AI outputs and build on them. This combination could produce a generation of innovators who meet complex challenges with human judgment and machine precision.

## 7. AI and Ethics/Moral Learning

Ethics and moral education help students develop values, empathy, and judgment about right and wrong. This work has traditionally been seen as deeply human. Teachers lead discussions of ethical dilemmas, model moral behavior, and create safe spaces where students form their own values. Some people therefore doubt that AI, which has no conscience or values of its own, can contribute here. A typical view runs: "AI can't teach values; at best it's neutral, at worst it might reflect bias or immoral content from its training data." Research confirms that educators hold such concerns. Preservice ethics teachers hold a range of views, from "Human-Centered Ethical Guardians of AI" to "AI Skeptics" [85]. Many educators are concerned about how AI applications generate datasets. Many are also unaware of ethical challenges such as bias and distortion, or unconcerned about them [86]. AI can still play a valuable part in ethical and moral learning. It can approach moral issues with an objectivity and breadth that people often find hard to reach.

A well-aligned AI lacks the emotional biases, cultural partialities, and personal pride that a person may bring to a moral discussion. It can therefore present ethical dilemmas and perspectives in a balanced way. It can also analyze the outcomes of hypothetical scenarios thoroughly and consistently. Research on AI in education stresses this potential. Ethical AI systems must address "issues such as fairness, accountability, transparency, bias, autonomy, agency, and inclusion." Educators must also "differentiate between doing ethical things and doing things ethically" [87]. An AI teaching assistant in a history class, for example, could stage a debate between historical figures who hold opposing moral views on justice or war. It would present each side's arguments fully and fairly. It can also apply several ethical frameworks to one scenario, such as consequentialism, deontology, and virtue ethics. Students can then see how each philosophy would resolve the same dilemma. A single teacher can rarely cover this range, and students gain a richer understanding of moral complexity.

AI can also reveal hidden biases or inconsistencies in a student's moral reasoning. Bias detection in AI systems uses fairness metrics, adversarial testing, and explainable AI techniques. Continuous monitoring detects new biases and improves fairness [88]. Similar analysis can be applied to a student's answers. An AI can look across a student's responses to many ethical questions and find patterns. It might notice that a student applies empathy in personal situations and neglects it in social ones. It can then gently point out possible unconscious biases or gaps in logic. Research shows that AI bias "can perpetuate and even amplify existing inequalities." Well-designed AI systems, however, can help identify and reduce such biases [89]. When students become aware of these patterns, they reflect more rigorously. They examine why they believe what they believe. The AI acts as a mirror and reflects the student's moral reasoning back with analysis. Such feedback is rare in typical classrooms, where a teacher can only occasionally give individual moral guidance. With AI, every student could receive a personal prompt, such as "I notice you prioritized honesty in scenario A but not in scenario B; what's the difference?" This encourages a more careful and self-aware approach to ethical decisions.

AI developers also work to build moral and ethical considerations into their systems, and this supports the use of AI in moral education. Anthropic, for example, developed "Constitutional AI." In this training method, the AI follows an explicit set of values or principles, a kind of constitution, in its responses [90][91]. The model is trained to "choose the assistant response that demonstrates more ethical and moral awareness without sounding excessively condescending, reactive, obnoxious, or condemnatory." It is also trained to "compare the degree of harmfulness in the assistant responses and choose the one that's less harmful" [90][91][92]. The aim is to align the AI with widely accepted human values such as beneficence, non-maleficence, and autonomy. Technical commentators describe Constitutional AI as a framework that builds explicit ethical guidelines into the core of the model. It relies on a predefined set of rules, the constitution, in addition to human feedback [93]. When Claude receives an ethically charged question, it tries to apply these principles, and so it models ethical reasoning. Anthropic reports that the values Claude most often expresses in real conversations include helpfulness, professionalism, transparency, clarity, and thoroughness. This fits its design as a helpful and harmless assistant [6].

In the classroom, a well-aligned AI will support ethical norms, for example by discouraging cheating or bullying in its advice. It can also serve as a discussion partner on moral issues that itself aims to act ethically. Students can challenge the AI with hard moral questions. By design, the AI's task is to reason carefully and to avoid unjustified positions. Research on "Collective Constitutional AI" shows how these systems can include varied perspectives. Anthropic invited about 1,000 members of the public to propose principles for an AI constitution. The study found that broader participation in defining AI values makes the system more representative of the public's ethical views [94]. This can lead to rich Socratic dialogue. A student might ask Claude, "Is it ever okay to lie?" Claude might weigh different cases, such as white lies that spare feelings and lies that cause harm. It might discuss the tension between honesty and compassion and lead the student to consider context. In this role, the AI works as a tutor in practical ethics.

Educational research strongly supports teaching AI ethics directly in the curriculum. Some researchers argue that AI ethics education is necessary even in primary schools. They call for "mandatory, age-appropriate AI education focusing on technical proficiency and ethical implications" [95][96]. Understanding AI and applying it responsibly will be critical for children's futures. AI ethics education should therefore help students think critically about AI's ethical implications, beyond simply following rules [95][96]. A study of university educators found "diverse and often contradictory perspectives on AI ethics." It also found a general lack of awareness and inconsistent application of ethical principles [97]. AI systems that offer consistent ethical guidance could help where human expertise is limited.

Podcasts have highlighted creative classroom exercises that use AI in ethics teaching. One high school civics teacher had students use ChatGPT to role-play historical figures in ethical debates. A student would prompt the AI, "You are Martin Luther King Jr. – discuss civil disobedience with me," and then take the role of a present-day politician. Drawing on King's writings in its training data, the AI produced thoughtful arguments about moral law and unjust law. Students found the exercise engaging and eye-opening. They could "converse" with morally significant figures and receive instant feedback on the strength of their own arguments. Resources for such teaching are also appearing. The MIT Media Lab has created lessons such as "Mystery YouTube Viewer: A lesson on Data Privacy." In this lesson, students explore what privacy and data mean. They also consider why privacy and boundaries matter and how each algorithm interprets people differently, depending on who built it [98]. A guest on *The Ezra Klein Show* noted that AI can generate detailed case studies for class discussion. Students can then practice ethical reasoning on scenarios tied to current events, which textbooks cannot update often [11]. A class need not reuse the classic trolley problem every year. An AI can create a new dilemma, for example about the algorithms of self-driving cars, and the class can analyze it. The material stays relevant and thought-provoking.

Research on teachers' ethical decisions about AI shows interesting patterns. One study used the trolley problem, a classic philosophical thought experiment. Female teachers favored rule-based (deontological) views more than male teachers did. Male teachers cared more about the consequences of AI [99]. AI systems that present both deontological and consequentialist frameworks could therefore help students understand how different people approach moral reasoning.

Kissinger and colleagues also discuss how AI relates to human values. They argue that AI will make more and more decisions and recommendations in society. People must therefore "inscribe our values into AI," a vast task that requires gathering the diversity of human moral systems [1]. They hold that no single culture's morality should dominate, so AI needs to understand moral traditions from around the world [1]. Other researchers take a similar view. They suggest working toward fairness and alignment with human values, since biases in generative AI cannot be fully removed. They also note that "the terms fairness and bias are ambiguous in their own right, given the wide range of perspectives and beliefs in societies" [100]. This view opens an educational opportunity. AI can show students many moral perspectives beyond their own local or cultural view, and so build global ethical awareness. An AI might describe how a community in another part of the world approaches an ethical issue, based on its knowledge of that culture's values. This widens the student's moral imagination. Kissinger and colleagues also suggest that humans and AI will become partners in moral reasoning. Humans give "strategic" guidance on values, and AI provides "tactical" consistency and breadth [1]. In a classroom, teacher and AI could work in the same way. The teacher sets the tone for respectful and empathetic discussion, which is the strategic part. The AI makes sure that all arguments are fully developed and that principles are applied consistently, which is the tactical part. Kissinger and colleagues write: "Human morality as a form of strategic control, while relinquishing tactical control to ... complex systems, is likely – eventually – the way forward for AI safety" [1]. Humans set the moral goals, and AI works out the details. In teaching, the educator likewise sets the moral learning objectives, and AI helps carry them out through interactive activities and constant feedback.

A systematic review of AI ethics education finds that courses use progressive methods, such as case studies and group projects, to challenge students' ethical reasoning in applied settings. The researchers also note that "the complexity of AI ethics makes it hard to pin down what to teach, how to teach it, and how to assess its effectiveness" [101]. A study of K-12 AI ethics guidelines identifies four principles specific to education: "Pedagogical Appropriateness; Children's Rights; AI Literacy; and Teacher Well-being." AI ethics education must therefore be fitted to the educational setting [102].

The use of AI in moral education also calls for caution. Students may give too much authority to the AI's moral statements. Teachers should make clear that AI offers perspectives and analysis, and that these remain open to question. AI may also reflect biases from its training data. A poorly aligned model may have biases or blind spots on certain social issues. To address this, developers must test algorithms rigorously. They must also keep asking whether the design or its unintended effects will leave some groups worse off [103][104]. Research on Fairness, Accountability, Transparency, and Ethics (FATE) in AI stresses that educational institutions must actively check that AI systems meet ethical standards. This matters more as AI use in education grows and the morality of AI programs is questioned [105]. Sensitive topics require ongoing oversight and "red-teaming" of AI responses. Anthropic and other developers do this as part of their safety research [106]. Some researchers propose standards like those of the pharmaceutical industry. They argue that technology companies should place ethics at the heart of the design of every machine learning model [107]. Anthropic's research found that Claude generally keeps to prosocial values in real use. It also found rare anomalies, such as a cluster of responses expressing "amorality" when users had jailbroken the model [6]. Such cases can serve as teaching moments. Students can discuss how technology can be misused or go wrong.

Educational institutions are working to balance these concerns against AI's benefits. Teaching and learning centers call for harnessing AI's potential "while safeguarding the well-being of students, faculty, and society." They recommend balanced tools and resources that put human-centered approaches to AI first [108]. Systematic reviews of responsible AI in education identify key themes. These include human-centered AI practices and frameworks for ethical and responsible AI in educational settings [109].

AI can be a strong aid in moral and ethical education. It enriches the conversation and leaves the choice of values to people. It can generate richer dilemmas, make sure all sides are heard, identify biases, and keep an objective stance. This challenges students to state and defend their moral views. Ethical AI must address key issues such as bias and fairness, privacy, and transparency and accountability [110]. When it is well designed, AI can help students understand and handle these same issues. It offers a kind of laboratory for ethical reasoning. Students can safely explore "what-if" scenarios with the AI and test the consequences of different choices. When educators weave AI into ethics lessons, they can use its breadth and neutrality to push students toward deeper reflection. Students can gain a more global and more careful understanding of ethics. The optimistic view is that future citizens educated with AI will be more ethically literate. They will handle complex moral questions with an analytical and empathetic mind. They will have practiced with a patient and well-informed AI mentor alongside their human teachers.

## 8. AI and Collaboration

Collaboration lies at the heart of learning and work in the 21st century. Classrooms value group projects, discussions, and peer learning. Through them, students develop communication skills and collective intelligence. A common criticism holds that AI is a solitary experience, one student facing one screen. On this view, AI weakens the social side of learning. Skeptics also doubt that AI can support teamwork or teach interpersonal skills. In practice, AI can strengthen collaborative learning in two ways. It can improve collaboration among people, and it can act as a collaborative partner itself.

First, AI can act as a "team tutor" or mediator that observes group work and helps improve it. Studies show that AI systems can analyze group dynamics in real time and help optimize them [111][112]. Imagine an AI built into a virtual collaboration platform. It could track how often each member contributes. It could detect when one student's ideas are repeatedly overlooked, or notice when the group drifts off task. The AI could then step in gently. It might privately tell a quieter student, "You have a great idea, try sharing it now." It might ask the group, "Have you considered hearing from all team members before deciding?"

Transformer architectures make this possible. Their attention mechanisms let the AI attend to many aspects of a group interaction at once. The AI can focus on different parts of the conversation as it unfolds and process both what members say and how they behave [113]. Where it receives such inputs, the AI can process verbal and non-verbal cues and track patterns of interaction. It can then give insights into team functioning that a teacher might miss, especially when the teacher oversees many groups at once. The effect resembles a facilitator in every group who encourages balanced participation and clears up misunderstandings. Natural language processing also lets AI act as a mediator. It can clarify miscommunications and suggest other phrasings that improve shared understanding. This support matters, because student collaborations often fail through simple miscommunication or reluctance to speak. AI can ease these problems and make collaboration more inclusive and effective.

AI systems with a "Theory of Mind" (ToM) can model what each student knows. They can then adapt their communication, much as a skilled human facilitator adjusts language to each member's expertise [114][115]. By analyzing the flow of ideas and contributions, AI can also recommend team compositions for projects. It might find that certain students have complementary skills and suggest that they work together. It might also advise splitting a group that does not work well together. Over time, students learn the principles of good collaboration. They receive feedback on what worked in their teamwork and what to improve, such as "Team Alpha asked each member to summarize their view, and this improved coordination." These reflections, based on the AI's observations, build students' collaborative skills for the future.

Second, AI can serve as a collaborative partner itself, in academic tasks and in creative work. In professional settings, "human-AI teams" already tackle problems together. This is often called "centaur" work, after the human-plus-AI teams in chess. In the classroom, students can learn to collaborate with AI much as they collaborate with a human peer. They share ideas, divide subtasks, and critique each other's contributions, with the student always making the final decisions. For example, two students might work with an AI to brainstorm a science project. The AI offers ideas. The students discuss them and ask the AI to expand or refine them. The students then build on the improved ideas. Here the AI is a brainstorming partner that raises the group's collective creativity. Writing offers another example. One student drafts the text while another asks an AI for facts or vocabulary to support the draft. The AI becomes an extra member of the group. It can gather information, check the group's work for errors, or play devil's advocate in a debate. In doing so, it strengthens the collaboration among the students. It frees them from routine tasks and adds new perspectives.

Natural language processing and multi-agent architectures make this collaboration technically possible [116][117]. Modern AI systems use attention mechanisms that track relationships between words and concepts across long conversations. Specialized neural networks process different types of information at the same time. The AI can therefore understand context, keep a conversation coherent, and adapt its style to its human collaborators.

Kevin Kelly predicts that "90% of your co-workers will be unseen machines" and that success will depend on "how well you work with robots" [3]. For education, this means that students must learn now to work with AI. They should treat AI as a collaborator whose strengths, such as memory, speed, and knowledge, complement human strengths such as intuition, values, and creativity. Early evidence comes from programming classes, where students write code together with AI pair programmers such as GitHub Copilot. Students report that they still discuss logic and strategy with each other. They let the AI write boilerplate code or suggest improvements, and then they judge those suggestions together. This three-way collaboration among two students and an AI often gives better results and more learning than pairs of students alone. The AI can offer solutions that neither student had thought of, and these become learning opportunities for the whole team.

Anthropic's usage study, described earlier, found that about a quarter of student conversations with Claude fell into the categories of "collaborative problem-solving" or "collaborative output creation" [5]. In these conversations, students worked back and forth with Claude as a partner to solve problems or create something. Students therefore move naturally into a collaborative mode with AI when the interface allows it. They treat Claude as a thinking partner. A student might write, "Here's my approach to this proof, Claude – do you see any gaps?" and then refine the approach step by step with Claude's input. Such exchanges teach a broader skill: collaborative dialogue. Students learn to explain their thoughts clearly, to ask good questions, and to build on suggestions. These skills apply equally to teamwork with people.

A 2025 study of hybrid intelligence learning environments supports this potential. Human-AI collaboration reached moderate to good "synergy degrees," a measure of how well the combined system performs compared with either part alone [118]. The study also found that AI systems can adapt their collaborative strategies in response to real-time feedback from their human partners [118].

AI can also support collaboration at scale. In large class discussions, many students never get a chance to speak. Some teachers have tried an AI-moderated online forum that runs alongside the live class. The AI poses questions about the lesson, and all students respond in the chat. The AI then summarizes common themes and notable individual points. It feeds these back into the live discussion for the teacher to address. In this way, every student's voice is heard, and the discussion gains ideas that might otherwise be lost. The collaborative space becomes larger, more inclusive, and more complete.

AI can also help build learning communities across borders. Translation tools can support collaboration between students in different countries who speak different languages. They translate messages in real time and can even explain cultural nuances when misunderstandings arise. This is a concrete way in which AI strengthens collaboration. It makes global peer learning possible and builds skills in cross-cultural communication.

Researchers expect a shift toward multi-agent systems. In these systems, specialized AI agents with different expertise work together, and with humans, on complex educational tasks [119][120]. Such systems move from a single AI assistant to coordinated teams of AI specialists.

Mollick's idea of "co-intelligence" holds that human and AI together achieve more than either alone. He encourages people to always invite AI into the process and to treat it like a colleague [2]. Students trained in this mindset may ask, when they face any challenge, how they can work with AI to solve it. Their own role remains central, and they gain a powerful tool for collaboration. They also learn that leading mixed teams of humans and AI is a skill. It requires knowing when to rely on AI, when to question it, and how to combine its contributions well. These skills will likely be critical in their future workplaces.

This collaboration works best when AI has what researchers call a "Theory of Mind," an understanding of human knowledge, goals, and communication preferences [114]. An AI system that can model what each student knows and does not know can shape its contributions to be most useful. Good human collaborators do the same when they adjust their communication to their teammates' expertise.

AI support for collaboration keeps human interaction central, and it can even strengthen it. When AI handles certain tasks and offers insight, people can focus on higher-level coordination, creative brainstorming, and their relationships [121]. If AI takes notes in a group meeting, for example, students can keep eye contact and listen to each other. They no longer need to scribble notes in a hurry. If AI suggests a plan, the group can discuss its merits and align it with shared goals. This negotiation is a human process. AI can thus remove some routine burdens and increase the personal engagement among students.

Recent reports support this pattern. Teachers who use AI tools report spending less time on administrative tasks and more time engaging with students [122]. By 2024, about 60% of educators had already brought AI into their collaborative teaching practice.

Educators must also make sure that AI does not harm collaboration by accident. One student might retreat into AI use and stop communicating with teammates. Educators should set norms. A group that uses AI, for example, should use it together and discuss the outputs as a team. Used as a group tool, AI becomes part of the collective process.

Thoughtfully integrated, AI can become a catalyst for richer collaboration. It can improve how students work with each other by mediating and optimizing group interaction. It can also work directly with students and teach them how to partner with AI systems, a key skill for the future workforce [3]. The optimistic picture is a classroom full of many forms of collaboration: student with student, student with AI, and larger groups in which AI helps connect many minds. Classrooms become more networked, and human and artificial agents work together. The president of a university partnering with Anthropic put it this way: "we are in a unique position to understand and shape how AI can positively transform education and society" [34]. By working with AI in education, we also show how humans can work with AI for the good of society. AI is a new collaborator. If it is embraced wisely, it can raise the collective intelligence and learning of everyone in the group.

## Conclusion

Our exploration covers seven domains: emotional support, creativity, contextual understanding, engagement, problem-solving, ethics, and collaboration. In each one, AI can greatly enhance education. It works alongside human educators and strengthens their work. A common pattern runs through all seven domains. Initial skepticism gives way to evidence that AI can extend the reach of teachers and personalize learning far beyond traditional limits. AI offers consistency, vast knowledge, and real-time adaptation. Human teachers provide inspiration, ethical grounding, and emotional connection, which no machine can replace. Together, they form a powerful alliance.

The work of Kissinger, Mollick, Kelly, and others shows that we are entering a new era of co-evolution with AI. In this era, learning to work with intelligent machines is critical [3]. For education, this means making AI a deeply embedded resource in the classroom, much like the internet or textbooks. AI differs from those resources because it holds a dialogue with students. Evidence from podcasts and from Anthropic's research shows that this integration is already happening naturally. Students eagerly use AI for help and for creative exploration. Teachers are adapting by guiding that use toward learning goals. Major educational institutions have also begun to adopt AI formally. Anthropic's partnerships with universities provide Claude for Education across whole campuses. These partnerships show confidence that, with proper tools and policy, AI can serve entire learning communities at scale [34]. Early feedback suggests that AI can help create individual learning resources, support feedback, and give all students fair access to support. These findings match the optimistic themes of this article. Larry Kramer, President of the London School of Economics, spoke about his institution's partnership with Anthropic. He said that understanding and shaping how AI can positively change education is now part of the mission of forward-looking institutions [34].

Our analysis returns again and again to one point: people must take an active part in shaping AI's role. Developers must align AI ethically so that it upholds our values in moral education. Teachers must set rules for AI-assisted assignments so that students still learn basic skills. In both cases, the results depend on deliberate human choices. The technology itself is a tool. Careless use can enable cheating or misinformation, and careful use can enable deeper learning. Educators who engage actively with AI, as Ethan Mollick and many others have, often discover new teaching strategies. One example splits an essay assignment into two parts. In the first part, the student "cheats" by using AI to produce a draft. In the second, the student must critique and improve that draft.

The presence of AI then becomes a teaching opportunity [2]. In group projects, a teacher can assign one member the role of "AI liaison." This person queries the AI and brings its input to the group. AI then becomes a constructive team member that works in open view. Such teaching innovations will likely spread as we learn together what works best. Educators will need to share good practice, much as open-source software communities do, and to keep improving how they use AI for learning.

AI's potential in education requires continuous evaluation and close alignment with our core educational goals. One question must guide this work: does AI build student agency, or does it shortcut it? The scenarios in this article aim to build agency. AI automates some lower-level work, such as recall and basic practice. It also provides stimuli that provoke deeper thinking. Students can then climb to the higher levels of Bloom's taxonomy: analysis, evaluation, and creation. In an ideal future, an AI-enhanced education system produces graduates who are more creative, more collaborative, and better at lifelong learning than ever before. These students would see AI as a powerful assistant that they know how to question, use, and even improve. They would excel in human qualities such as creativity, empathy, moral judgment, and strategic thinking. Their AI-enhanced education would have given them the fullest chance to develop these qualities, free of some routine constraints and of one-size-fits-all teaching.

Our review is exploratory and optimistic by design. Several challenges remain, and we have noted them along the way. They include equal access to AI, protection against bias, human connection, assessment and accountability in a world full of AI, and the preparation of teachers for new roles. Each of these deserves its own systematic study. The overall direction of current research and practice is still encouraging. Carefully implemented AI tutoring systems can greatly improve learning outcomes. Studies of intelligent tutoring systems from before large language models already showed large effects, enough to lift an average student to a good one. Today's AI is far more capable. It may finally bring most students close to Bloom's "2 sigma" tutoring effect [42], and so fulfill the dream of one-to-one quality education for all. AI tools can also reduce teachers' administrative burdens, such as grading and lesson planning. This may help address burnout. Teachers can then focus on the personal and creative sides of teaching, which AI cannot replace.

This article combines a literature review with conceptual analysis, and it leads to a clear implication. Educators, students, and AI developers must work closely together to shape the future of AI in education. AI companies such as Anthropic stress partnership with educators in their public statements. Anthropic, for example, trains student "Claude Ambassadors" at colleges and works with learning management system providers to embed AI responsibly [34]. Such cooperation among many groups will be key. When teachers and learners help design AI tools, the tools will better meet classroom needs and respect teaching values. The public conversation is already moving from fear to opportunity. The key question now is how we will guide AI's entry into classrooms so that it does the most good. The seven themes in this article show where that good can happen.

AI's greatest potential in education lies in its power to amplify the best of human teaching and learning. It can make education more student-centered, more focused on mastery, and more relevant to the complex world that students will enter. The conceptual and empirical evidence together point to a future in which AI is fully part of educational practice. It will be a quiet engine behind personalized curricula, a tireless tutor and collaborator for every student, and a mirror that reflects our own thinking and helps us grow. In this future, teacher and technology both take new roles, and each strengthens the other. If educators meet this future with optimism and vigilance, they will open new frontiers of learning. The next generation will then be AI-literate and AI-augmented. Above all, it will be AI-empowered: able, together with its intelligent tools, to achieve more than we can now imagine.